\documentclass[prd,twocolumn]{revtex4}
\usepackage{graphicx}
\begin{document}

\title{Removal of $Z_3$-symmetry breaking from Fermionic Determinants}

\author{Toru T. Takahashi}

\affiliation{Yukawa Institute for Theoretical Physics, Kyoto University,
Kitashirakawa-Oiwakecho, Sakyo, Kyoto 606-8502, Japan}

\date{\today}

\begin{abstract}
We consider prescriptions that 
are free from the direct charge-screening effects by quark loops
and enable us to clarify the confining nature of a vacuum.
We test two candidates for an order parameter,
a Polyakov loop ($P$) evaluated in zero-triality backgrounds
and fermionic determinants (${\mathcal D}_{1,2}$)
with non-zero triality.
Especially, ${\mathcal D}_{1,2}$
has very small fluctuations in comparison with a Polyakov loop
in zero-triality sector,
and seems to well reflect the characteristic of a vacuum.
Such prescriptions could be still usable for the clarification
of the confinement property of a vacuum.
\end{abstract}

\maketitle

\section{Introduction}

Clarification of the chiral phase transition 
and the confinement/deconfinement transition
are the longstanding central issues in QCD.
There have been numerous efforts to understand
the phase structure of QCD at zero and finite temperature or density 
regions~\cite{Philipsen:2007rj}.
Chiral phase transition is detected by monitoring
the order parameter, chiral condensate $\langle\bar\psi \psi\rangle$,
whereas a Polyakov loop is widely used
in order to know whether a system is in the confinement phase or not.
The expectation value of a Polyakov loop is zero
when a system is in the confinement phase,
and it can be finite in the deconfinement phase.
A Polyakov loop may be considered as an order parameter for the 
color confinement.
These two ``order parameters'' unfortunately do not go together.
(Interestingly, a Polyakov loop and Dirac eigenvalues,
which also reflect the chiral phase transition,
can be related~\cite{Gattringer:2006ci,Bruckmann:2006kx}.)
While chiral condensate $\langle \bar \psi \psi \rangle$
can be a good order parameter
if we neglect the small current quark masses,
a Polyakov loop can serve as an order parameter
only in pure gauge systems without quarks (or with infinitely heavy quarks),
where deconfinement transition is expressed by
the spontaneous $Z_3$-symmetry breaking.
The presence of dynamical quarks explicitly breaks the $Z_3$ symmetry
and a Polyakov loop results in an approximate order parameter.
The color fields induced by a Polyakov loop
is readily screened by the dynamical quark loops
that twist along the Euclidean time direction.
Nevertheless, a Polyakov loop surely reflects the confinement nature
and has been used in many studies.

It may be however desired 
to clarify the confinement nature in a clearer manner.
For example, the chiral symmetry breaking surely has an impact
on the properties of quarks.
Current quarks would change their natures due to the spontaneous 
chiral symmetry breaking acquiring the large effective masses
~\cite{Nambu:1961tp,Nambu:1961fr},
and therefore a Polyakov loop screened by ``constituent'' quarks
could be affected by the chiral phase transition.
The apparent coincidence
of the critical temperatures for the chiral and confinement
transitions is still under debate.
Even if we evaluate a Polyakov loop correlator
$\langle P(0)P(x) \rangle$,
it is not so straightforward to clarify the color-confinement nature
since we always have string-breaking effects.
In high-density systems, the concept of color confinement could be obscure.
Quarks may freely move from one hadron to another 
in sufficiently dense hadronic systems,
even if the vacuum is still in ``color confinement'' phase.
In fact, ``quarkyonic'' phase has been proposed
recently~\cite{McLerran:2007qj,Miura:2008gd},
where quarks are confined but their degrees of freedom 
can dominate the system.
In any case, as long as we insist on a Polyakov loop,
it is needed to single out the vacuum property
independently of the direct charge-screening effects.
Search for order parameters have a long-standing history,
and several studies have been performed so far
~\cite{Detar:1982wp,Fukushima:2002bk,Faber:1995up,Kratochvila:2006jx}.
We consider prescriptions that remove the dynamical quark loops
which directly screen the color fields from a Polyakov loop.

\section{Fermionic Determinant}

We introduce the basic ideas
~\cite{Detar:1982wp,Fukushima:2002bk,Faber:1995up,Kratochvila:2006jx}
in this section to make this paper self-contained.
We assume SU(3) lattice gauge theory with single quark-flavor 
in what follows. (Extensions to other cases are simple.)
The QCD partition function is expressed as
\begin{eqnarray}
{\mathcal Z}
=
\int dU \det D\ 
e^{-S_g[U]},
\end{eqnarray}
with $D$ a (lattice) Dirac operator,
in which non-zero chemical potential can be introduced,
and $S_g[U]$ being the gauge action.
In advance of the gauge fields' integration,
quark fields are integrated out,
and all the quark dynamics is encoded in $\det D$.
The quark loops that can directly screen the color fields
from a Polyakov loop originate from such a fermionic determinant.

\begin{figure}[h]
\includegraphics[scale=0.27]{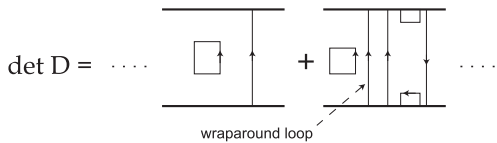}
\caption{\label{loops}
Schematic figure for the expansion of a fermionic determinant
in terms of quark loops.
}
\end{figure}

This fermionic determinant 
can be expanded in terms of two types of quark loops (See Fig.~\ref{loops});
ordinary quark loops and wraparound quark loops.
Ordinary loops exist entirely in a system and can be smoothly shrunk.
Wraparound loops twist along the imaginary-time direction,
and they are the very loops that can directly screen a Polyakov loop.
Such wraparound loops are also responsible for the finiteness
of quark density at finite chemical potential,
and are important ingredients.

A fermionic determinant can be generally divided into three terms;
\begin{equation}
{\mathcal D}
\equiv
\det D 
=
\sum_{i=0,1,2} {\mathcal D}_i
\end{equation}
with ${\mathcal D}_i$ ($i=0,1,2$) the terms which contain
$(3k+i)$ wraparound quark loops, respectively.
Here $k$ is an integer and can be negative,
and a wraparound anti-quark loop is counted as $-1$.
One connected quark loop which twists $m$-times along the 
Euclidean time direction is counted as $m$ quark loops.
The total number of quark loops is defined as a net number:
In case a contribution contains as many wraparound quark loops 
as wraparound anti-quark loops, the number is defined as zero.
We concentrate only on wraparound loops, since
ordinary quark loops are irrelevant in the following argument.
Next, we consider the uniform $Z_3$ transformation;
$U_t(t=0)\rightarrow zU_t(t=0)$ with $z\equiv \exp(2/3\pi i)$.
Here $U_t(t=0)$ are the link variables on a lattice.
By such a transformation,
a Polyakov loop $P$ is rotated as $P\rightarrow zP$.
This $Z_3$ transformation also affects
dynamical (wraparound) quark loops, and ${\mathcal D}_i$
are transformed as ${\mathcal D}_i\rightarrow z^i {\mathcal D}_i$.
Using this property, we can single out each ${\mathcal D}_i$
by means of such $Z_3$ link-variable transformations.
Denoting the fermionic determinant 
after $n$-times such $Z_3$ transformations
as ${\mathcal D}(n)$, ${\mathcal D}_i$ can be obtained as
\begin{eqnarray}
{\mathcal D}_i = \frac13 \sum_{n=0,1,2}{\mathcal D}(n)
\times
z^{-in}.
\end{eqnarray}
(See earlier works~\cite{Detar:1982wp,Fukushima:2002bk,Faber:1995up,Kratochvila:2006jx}.)
With vanishing quark chemical potential,
${\mathcal D}_0^* = {\mathcal D}_0$ and ${\mathcal D}_1^* = {\mathcal D}_2$ hold
due to the charge conjugation symmetry.
As we have mentioned above,
${\mathcal D}_i$ contains $(3k+i)$ wraparound quark loops,
and hence what screens the Polyakov loop's color fields are
the quarks encoded in ${\mathcal D}_2$.
In fact, the complex phase associated with ${\mathcal D}_2$
rotates a Polyakov loop into the real sector.
(${\mathcal D}_0$ and ${\mathcal D}_1$ do not.)
The important note is that
quarks in ${\mathcal D}_0$ and ${\mathcal D}_1$
cannot screen a fundamental charge completely.
Then, if we evaluate only ${\mathcal D}_0$,
the response of a Polyakov loop would be the same
as that obtained in a quenched system.
${\mathcal D}_0$ is in fact $Z_3$-transformation invariant
and does not cause the explicit $Z_3$-symmetry breaking.
We note here that this removal of the explicit $Z_3$-symmetry breaking
is essentially different from the quark-loop quenching.
Evaluating ${\mathcal D}_i$
corresponds to computing the partition function ${\mathcal Z}_i$
in the $i$-triality sector (${\mathcal T}=i$),
\begin{equation}
{\mathcal Z}_i
=
\sum_{|3k+i\rangle}
\langle 3k+i|
e^{-\beta \hat H}\ 
|3k+i\rangle.
\end{equation}
Here, $|3k+i\rangle$ denote
all the possible states that contain $(3k+i)$ net quarks
($k$ is an arbitrary integer).
The vacuum still contains dynamical quark loops,
and if the quark chemical potential is finite,
there will exist $3k$ net quarks ($k$ net ``baryons'') in the system.
String-breaking phenomena caused by dynamical quarks 
will be also reproduced.
It was also shown in Ref.~\cite{Kratochvila:2006jx}
that such a projection does not change thermal properties.
We expect that 
it would be possible 
to unambiguously clarify the confining property of a vacuum
independently of the direct charge-screening effect by dynamical quarks.
(If we employ an anti Polyakov loop, 
the roles of ${\mathcal D}_1$ and ${\mathcal D}_2$ 
switch positions with each other.)

\section{Numerical tests}

\subsection{Polyakov loop}

In this subsection, we individually and explicitly
investigate the distributions of a Polyakov loop
evaluated with ${\mathcal D}_i(i=0,1,2)$.
We generate quenched gauge configurations with the plaquette action
at $\beta =5.7$ on $4^4$ lattice,
and perform reweighting using ${\mathcal D}_i$
as well as the full determinant ${\mathcal D}$
generated with the Wilson fermion at $\kappa =0.1600$.

\begin{figure}[h]
\includegraphics[scale=0.27,origin=c,angle=-90]{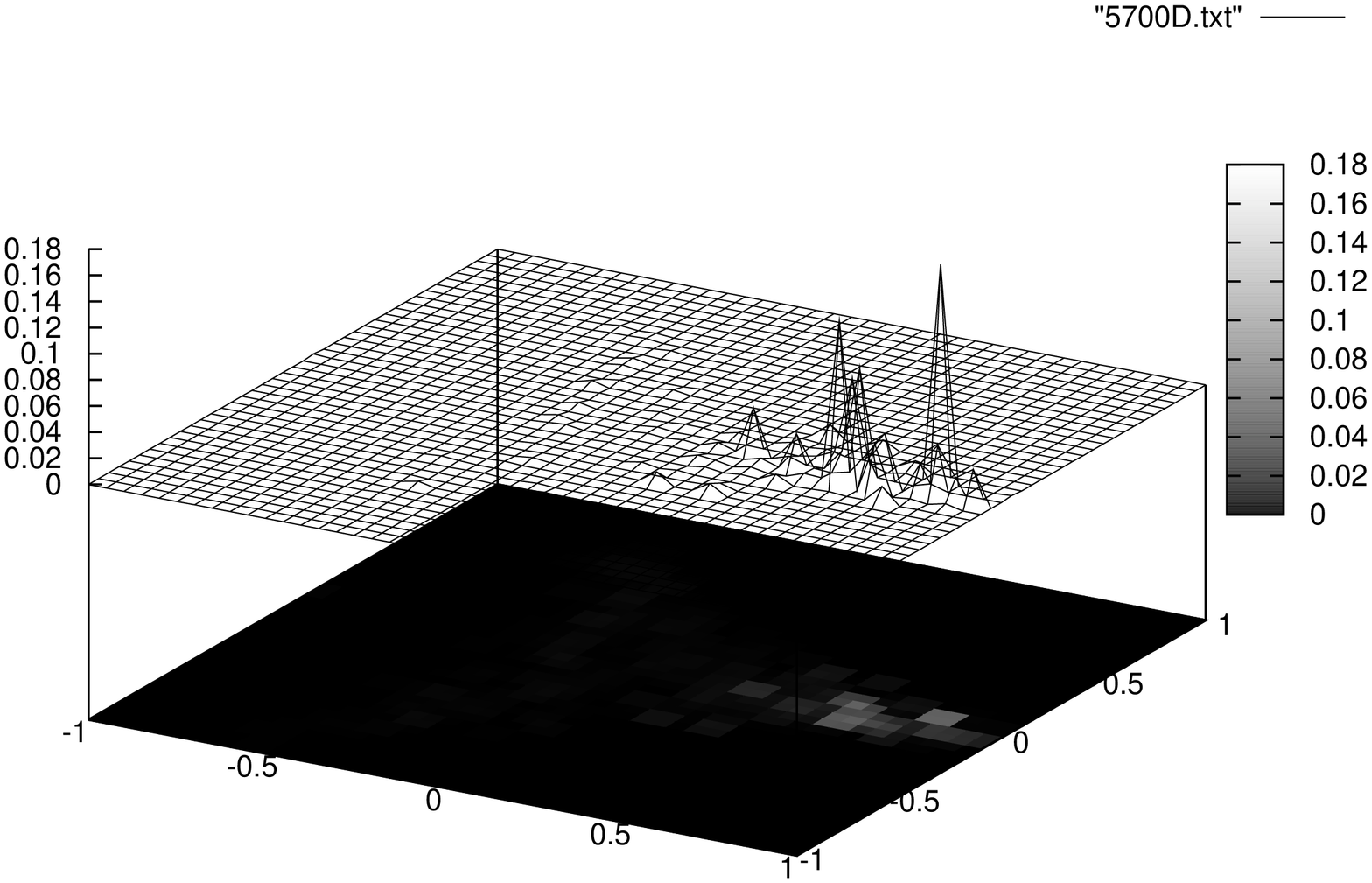}
\includegraphics[scale=0.27,origin=c,angle=-90]{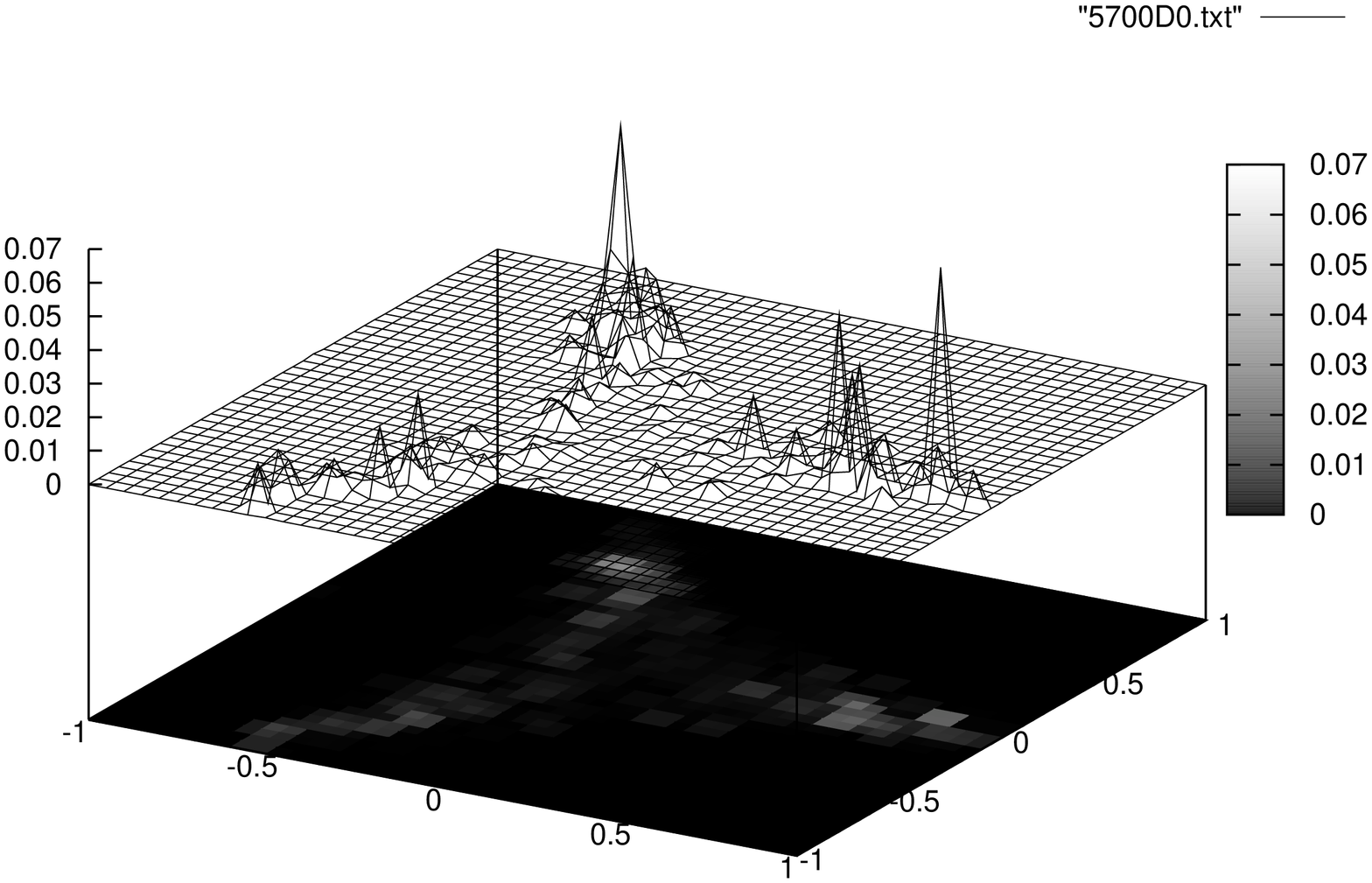}
\caption{\label{z3bk}
The distributions of the Polyakov loops 
evaluated with ${\mathcal D}$(upper) and ${\mathcal D}_0$(lower)
at $\beta=5.7$, where the broken $Z_3$ symmetry is observed.
}
\end{figure}

\begin{figure}[h]
\includegraphics[scale=0.27,origin=c,angle=-90]{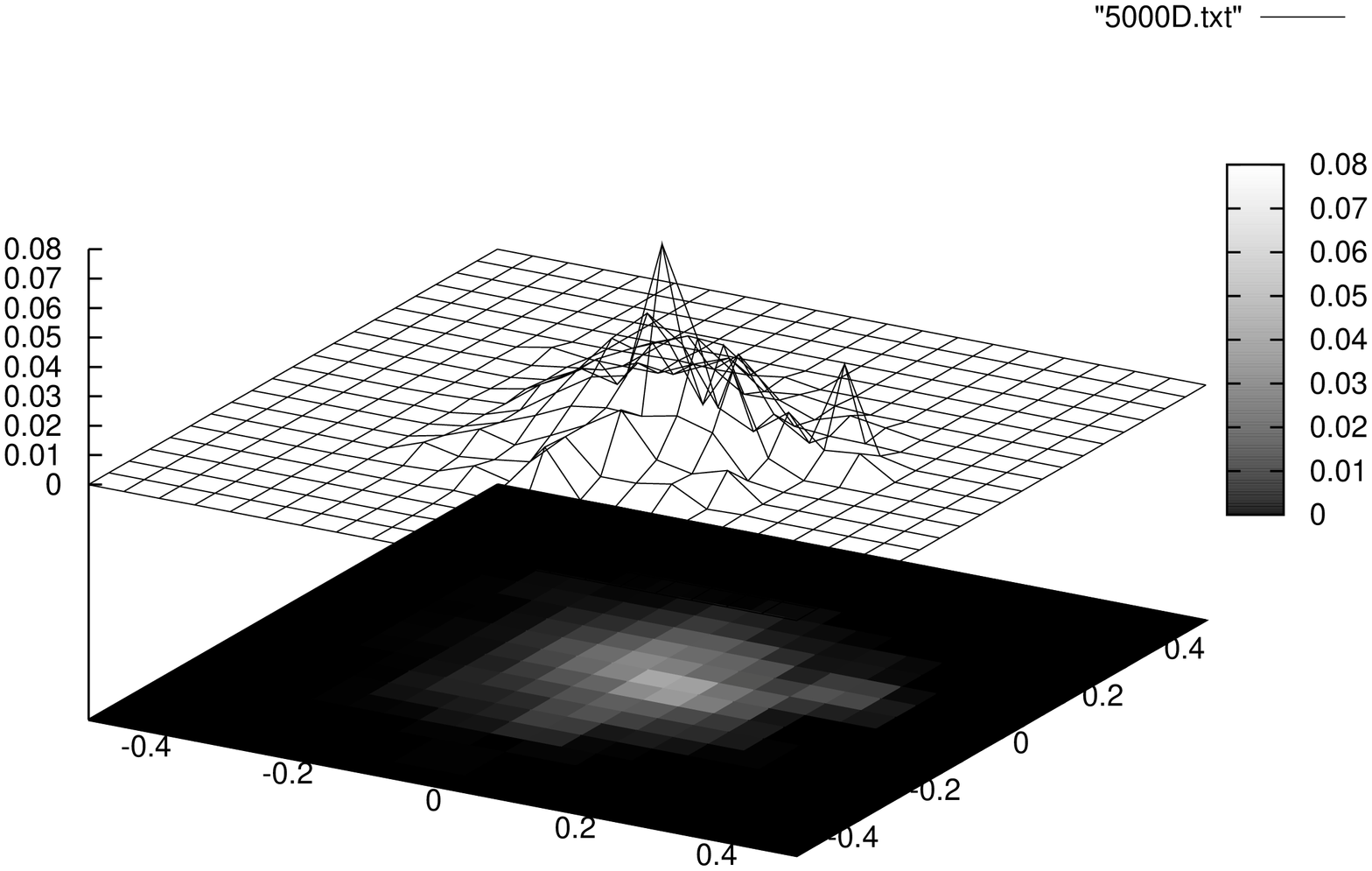}
\includegraphics[scale=0.27,origin=c,angle=-90]{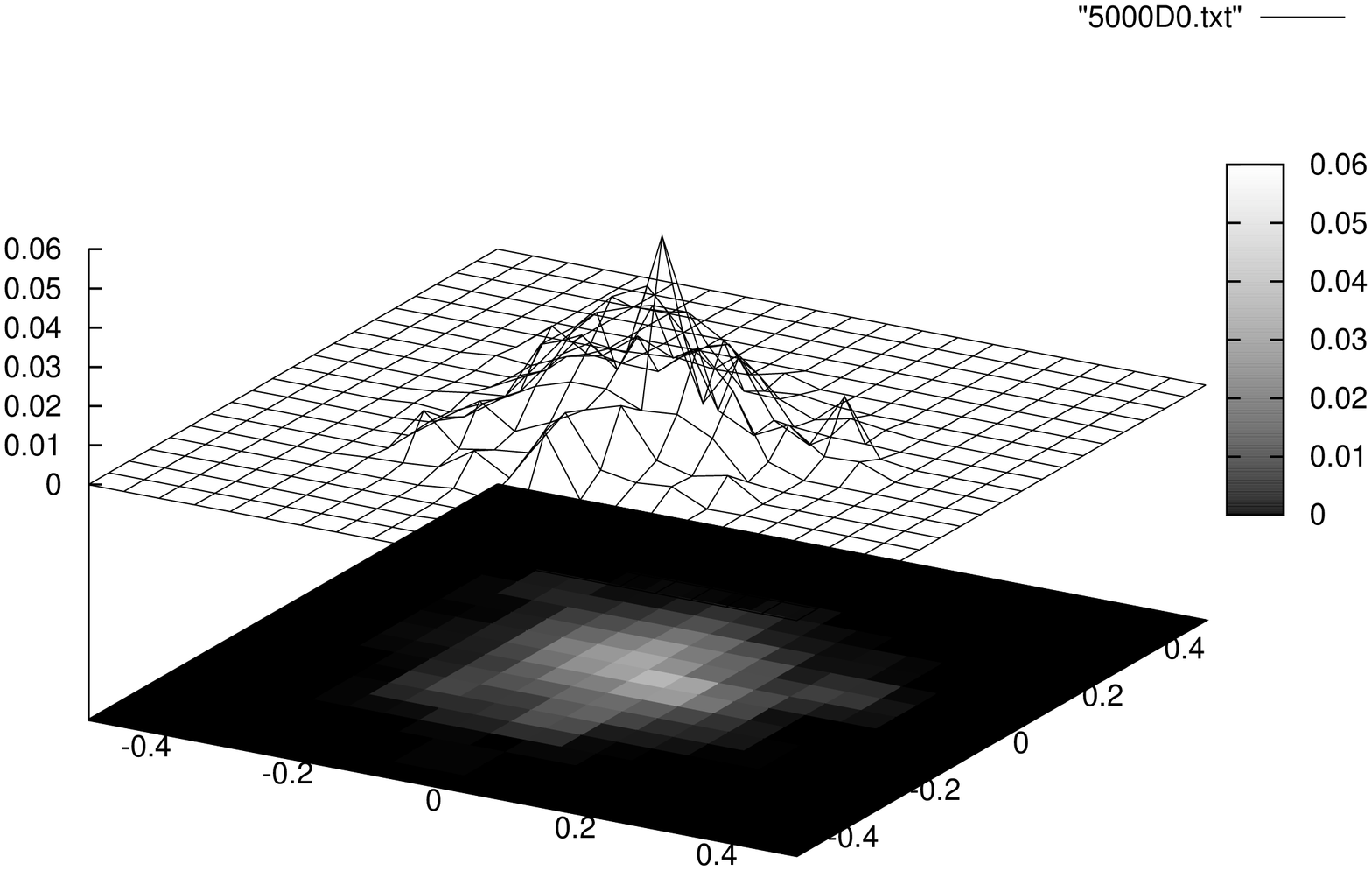}
\caption{\label{z3ok}
The distributions of the Polyakov loops
evaluated with ${\mathcal D}$(upper) and ${\mathcal D}_0$(lower)
at $\beta=5.0$, where no $Z_3$ symmetry breaking is observed.
}
\end{figure}

Fig.~\ref{z3bk}(upper) and Fig.~\ref{z3bk}(lower)
are the Polyakov-loop distributions in the complex plain
evaluated with ${\mathcal D}$ and ${\mathcal D}_0$, respectively,
at $\beta=5.7$, where the broken $Z_3$ symmetry is observed.
As expected, the strength in the real sector
is much enhanced by the effect of ${\mathcal D}$,
which can be seen in Fig.~\ref{z3bk}(upper).
On the other hand, Fig.~\ref{z3bk}(lower) shows
the similar strength to the quenched case,
which was expected in the previous section.
Especially one can find a three-peak structure in the complex plain.

We show the same plots obtained at $\beta=5.0$
in Fig.~\ref{z3ok}, where no $Z_3$ breaking can be found.
The strength in the real sector is again (but weakly) enhanced 
by the effect of ${\mathcal D}$ as can be seen in Fig.~\ref{z3ok}(upper),
whereas Fig.~\ref{z3ok}(lower) shows
the similar distributions to the quenched system.
This prescription, 
where we measure a Polyakov loop in zero-triality backgrounds 
(evaluating ${\mathcal D}_0$),
is free from the direct screening effects or
the string-breaking effects by quark loops,
and seems to enable us to clarify the confining nature of a vacuum.

\begin{figure}[h]
\includegraphics[scale=0.27,origin=c,angle=-90]{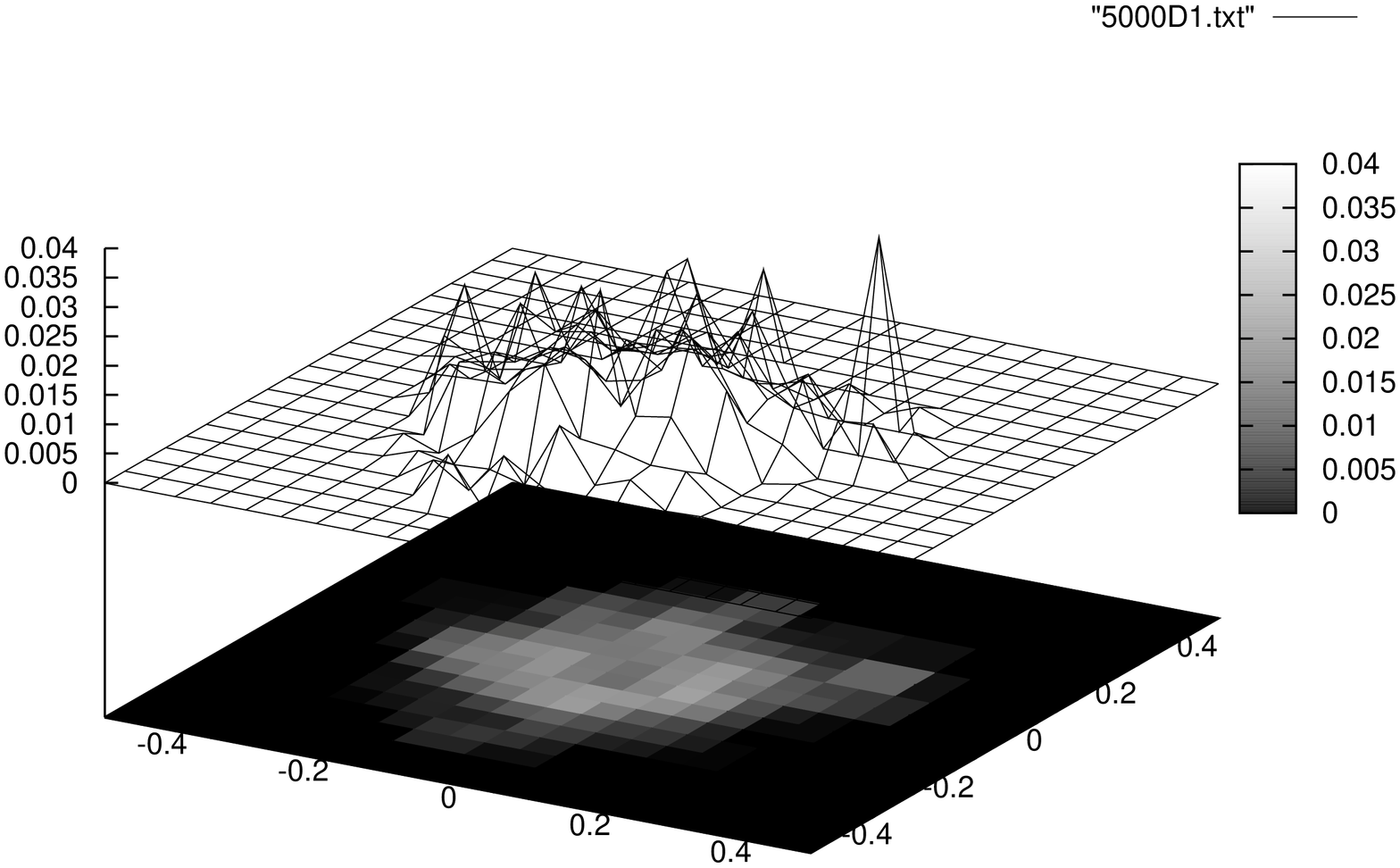}
\includegraphics[scale=0.27,origin=c,angle=-90]{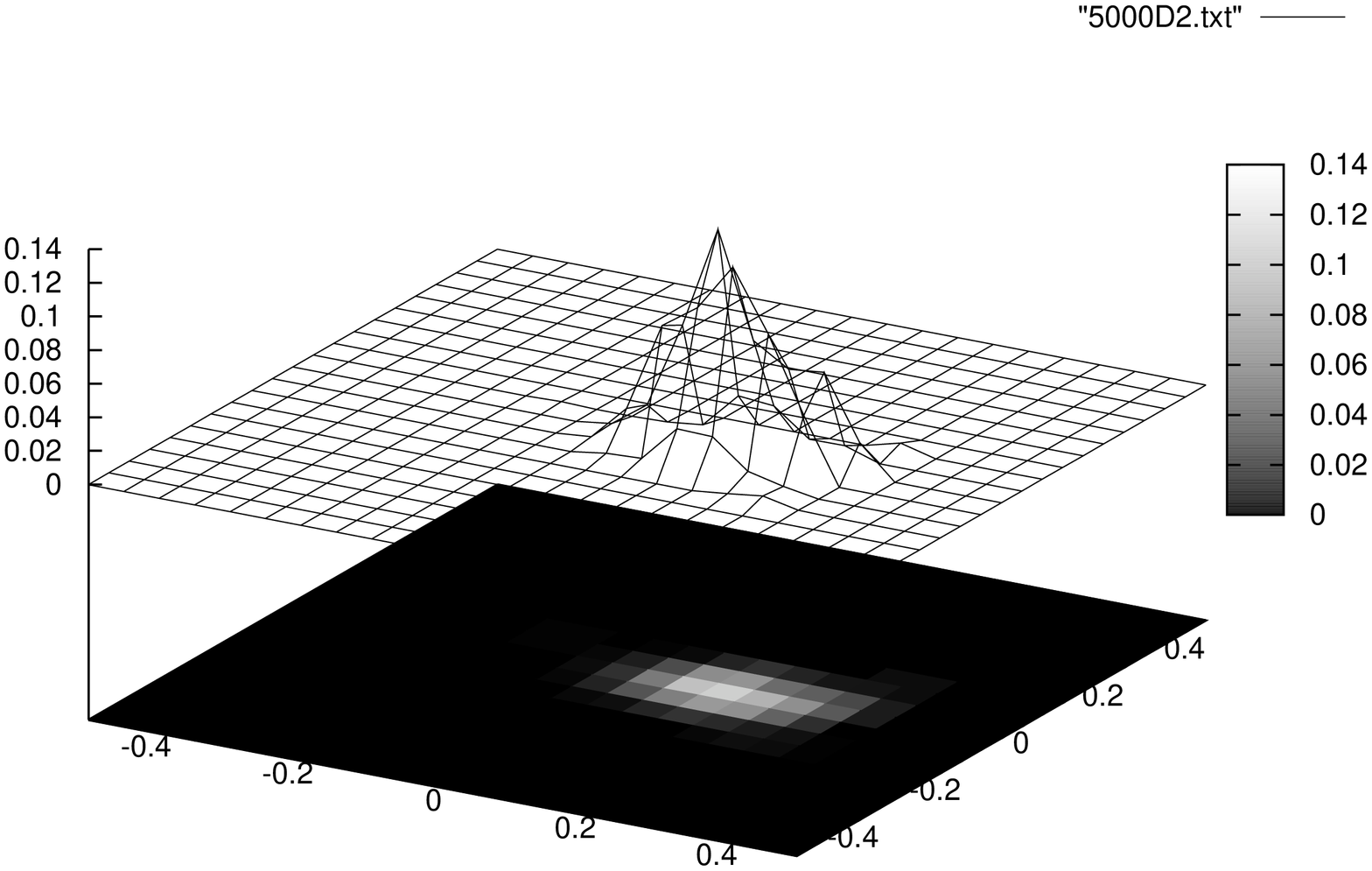}
\caption{\label{d1d2}
The distributions of the Polyakov loops
evaluated with ${\mathcal D}_1$(upper) and ${\mathcal D}_2$(lower)
at $\beta =5.0$.
}
\end{figure}

For further clarification, we present the distributions
reweighted with ${\mathcal D}_{1,2}$ in Fig.~\ref{d1d2}.
One can observe that 
only ${\mathcal D}_2$ can rotate a Polyakov loop into the real sector,
which implies that directly-screening quarks are coming from ${\mathcal D}_2$.
The Polyakov loops evaluated with ${\mathcal D}_1$
are rather uniformly distributed around the origin.
Then, ${\mathcal D}_1$ or ${\mathcal D}_0+{\mathcal D}_1$
may be also employed for this prescription.
The nonvanishing Polyakov loops
are sometimes related to the explicit $Z_3$-symmetry breaking
caused by fermions.
Including ${\mathcal D}_1$ actually breaks the $Z_3$ symmetry explicitly,
but it gives a similar Polyakov loop distribution
to the $Z_3$-symmetric quenched case,
which is natural
if we take into account that quarks in ${\mathcal D}_1$ cannot screen
a Polyakov loop completely and hence the situation
is physically similar to the quenched system.

So far, the response of the Polyakov loop
evaluated with ${\mathcal D}_0$ has been simply similar to the quenched case,
and the difference is unclear.
In order to show the difference,
we perform a test with non-vanishing quark chemical potentials.
We hereafter focus on the phase angles of ${\mathcal D}_0$ and ${\mathcal D}$
at finite chemical potential.
(Averaging the fermionic determinants at finite chemical potential
was suggested and performed some years ago in the context of
the reduction of phase fluctuations~\cite{deForcrand:1999cy,Aarts:2001dz,Sasai:2003py}.)

\begin{figure}[h]
\includegraphics[scale=0.27]{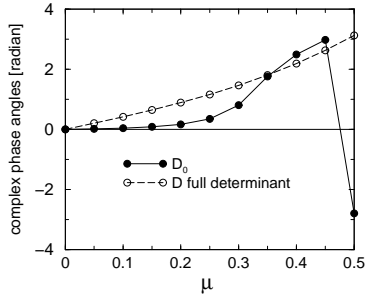}
\caption{\label{phases}
The phase angles of ${\mathcal D}_0$ and ${\mathcal D}$
obtained as a function of chemical potential $\mu$ are plotted.
They were evaluated with one gauge configuration.
}
\end{figure}

Fig.~\ref{phases} shows
the phase angles of ${\mathcal D}_0$ and ${\mathcal D}$
obtained at $\beta=5.7$, $\kappa =0.1640$, and $0\le \mu \le 0.5$,
with one gauge configuration.
The phase angle of the full determinant ${\mathcal D}$
grows as we increase the chemical potential $\mu$,
which is caused by the asymmetry between quarks and antiquarks.
However, as can be found in Fig.~\ref{phases},
the phase angle of ${\mathcal D}_0$ remains zero in the small-$\mu$ region,
and rapidly grows above some value of $\mu$.
The reason for this behavior would be that
the number of wraparound quarks in ${\mathcal D}_0$ is $3k$.
When $\mu$ is small enough, 
it is not likely for the system to accommodate three net quarks,
and ${\mathcal D}_0$ is still symmetric between quarks and anti-quarks,
which implies $k\sim 0$.
The phase angle at small-$\mu$ region is brought about
by ${\mathcal D}_1$ and ${\mathcal D}_2$.
At large $\mu$, the charge conjugation symmetry in ${\mathcal D}_0$
is largely broken, and the phase angle can be finite, which implies $k>0$.

\subsection{Distributions of Fermionic determinants}

\begin{figure}[h]
\includegraphics[scale=0.27]{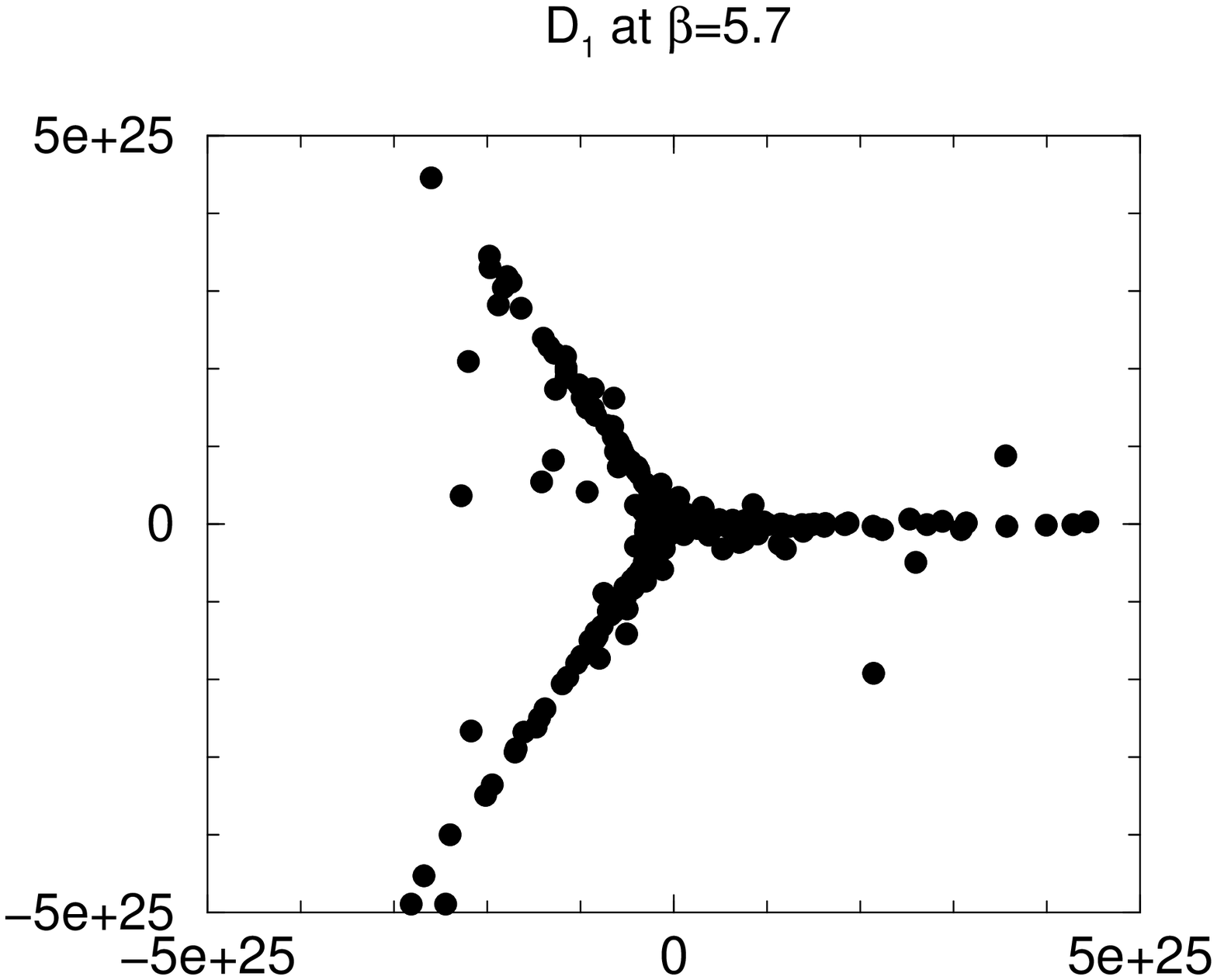}
\includegraphics[scale=0.27]{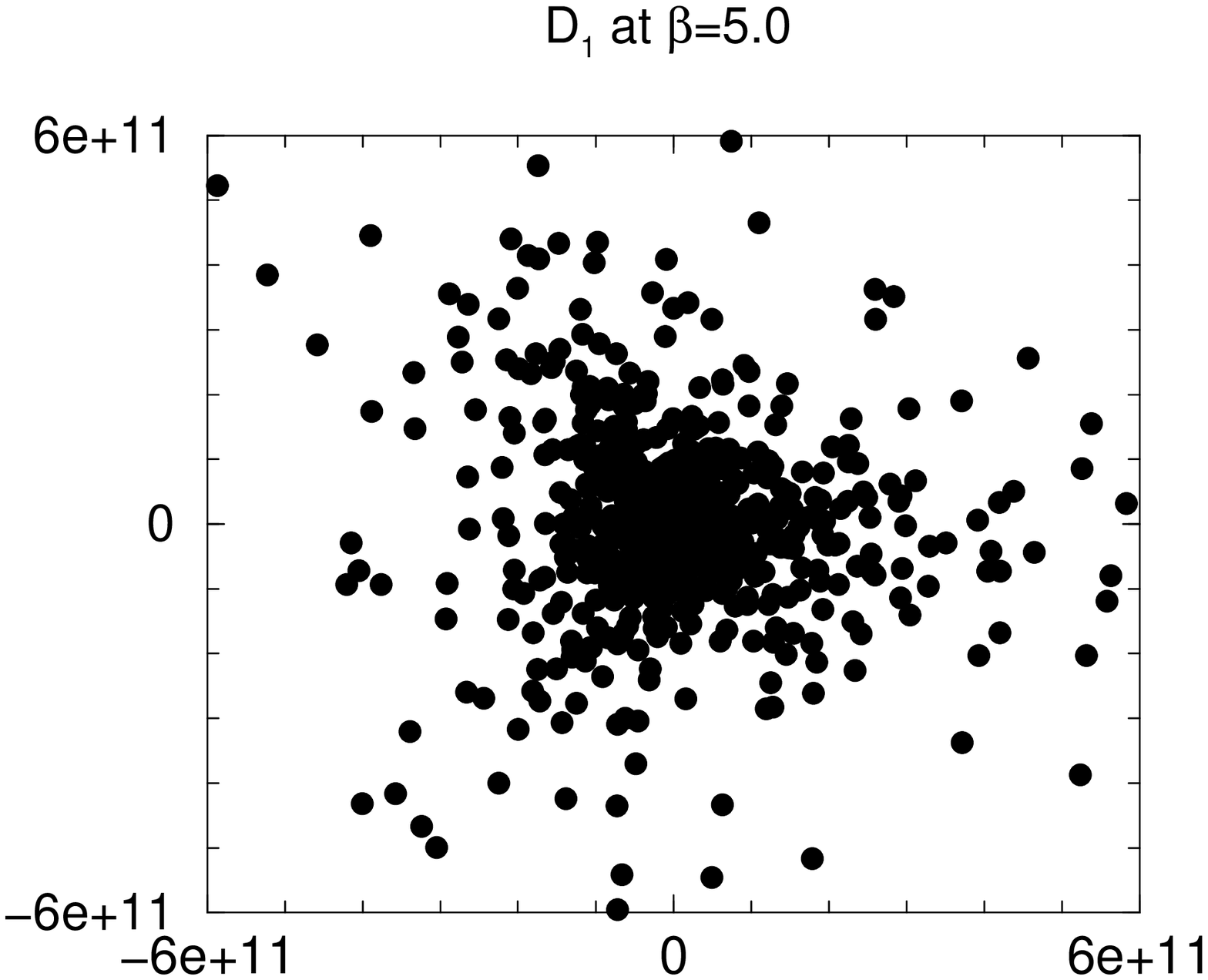}
\caption{\label{distFD}
The distributions of ${\mathcal D}_1$
computed at $\beta = 5.7$(upper) and $\beta = 5.0$(lower) are plotted
in the complex plain.
These distributions 
are qualitatively similar to the Polyakov-loop distributions
evaluated with ${\mathcal D}_0$.
}
\end{figure}

We next take a look of the distributions 
of fermionic determinants themselves 
in quenched QCD in this subsection.
Fig.~\ref{distFD} shows the distributions of ${\mathcal D}_1$
in the complex plain
computed at $\beta = 5.7$(upper) and $\beta = 5.0$(lower).
These distributions 
are qualitatively similar to the Polyakov-loop distributions
evaluated with ${\mathcal D}_0$.
Such behaviors can be understood intuitively:
${\mathcal D}_1$ contains $(3k+1)$ wraparound quark loops,
and hence corresponds to the free energy 
of $(3k+1)$ light quarks~\cite{Detar:1982wp}.
The physical situation is similar to the $3k$-quark vacuum 
(obtained by evaluating ${\mathcal D}_0$)
with one Polyakov loop,
where $(3k+1)$ net charges exist.
${\mathcal D}_{1,2}$ can be another candidate
for an order parameter~\cite{Detar:1982wp}.

If we directly generate 
unquenched gauge configurations adopting ${\mathcal D}_0$ instead of 
${\mathcal D}$
via some adequate algorithms~\cite{Kratochvila:2006jx,Alexandru:2005ix},
what is measured is the ratio,
${\mathcal D}_{1,2} / {\mathcal D}_0$,
which will yields ${\mathcal O}(1)$ values.
${\mathcal D}_{1,2} / {\mathcal D}_0$ measured in zero-triality sector
gives
\begin{equation}
{\langle {\mathcal D}_{n} / {\mathcal D}_0\rangle^{{\mathcal T}=0}}
=
\frac
{\int DU {\mathcal D}_{n} \ e^{-S_g[U]}}
{\int DU {\mathcal D}_{0} \ e^{-S_g[U]}},
\end{equation}
which differs in normalization from the original proposal 
in Ref.~\cite{Detar:1982wp},
\begin{equation}
\frac
{\int DU {\mathcal D}_{n} \ e^{-S_g[U]}}
{\int DU {\mathcal D}     \ e^{-S_g[U]}}.
\label{DMOP}
\end{equation}
The latter can be obtained by measuring
\begin{equation}
\frac{\langle {\mathcal D}_{n} / {\mathcal D}_0\rangle^{{\mathcal T}=0}}
{\langle {\mathcal D} / {\mathcal D}_0 \rangle^{{\mathcal T}=0}}
=
\frac
{\int DU {\mathcal D}_{n} \ e^{-S_g[U]}}
{\int DU {\mathcal D}     \ e^{-S_g[U]}},
\end{equation}
in zero-triality sector.
${\mathcal D}_{1,2} / {\mathcal D}_0$ 
as well as a Polyakov loop
evaluated in zero-triality sector
also seems usable for the clarification of the color confinement.

If we allow non-zero triality sectors 
in unquenched gauge-updation processes (ordinary updations),
the order parameter should be ${\mathcal D}_{1,2} / {\mathcal D}$,
which also turns out to be 
the original form proposed in Ref.~\cite{Detar:1982wp} (Eq.(\ref{DMOP})).
This implementation, however, would not be suitable for such purposes,
simply because ergodicity is worse.
Though the configurations that have smaller statistical weights 
${\mathcal D}e^{-S_g}$ will produce larger ${\mathcal D}_{1,2} / {\mathcal D}$
and $Z_3$ symmetry 
in $\langle {\mathcal D}_{1,2} / {\mathcal D}\rangle$
could be finally recovered,
such configurations would less appear in actual calculations.

We note here that the inevitable failure of such treatments 
in the thermodynamic limit
was discussed in Ref.~\cite{Fukushima:2002bk}.
Even so, if we directly measure 
$\langle {\mathcal D}_n \rangle / \langle {\mathcal D} \rangle$
in quenched backgrounds
(like our analyses in this paper 
and the original form in Ref.~\cite{Detar:1982wp}), 
the ergodicity itself is not lost~\cite{Fukushima:2002bk}
and this simple prescription could be valid,
though it is time-consuming.

\subsection{Spontaneous $Z_3$-symmetry breaking}

The deconfinement phase transition, if it exists, would be detected
still as the spontaneous $Z_3$-symmetry breaking, as in the quenched case.
(Even when the transition is crossover, the argument is not invalidated.)
We have to introduce a small explicit $Z_3$-symmetry breaking effect
and take the thermodynamic limit before removing the explicit breaking,
in finite volume systems.
Practically, it will be workable for the detection of transition
to evaluate $({\rm Re}\Omega^3)^{\frac13}$.
Here $\Omega$ is a Polyakov loop $P$
or the fermionic determinants ${\mathcal D}_{1,2}$.
Especially, the fluctuation of ${\mathcal D}_{1,2}$
is much suppressed than that of a Polyakov loop in zero-triality sector,
which can be seen in Fig.~\ref{distFD}(upper):
The values of ${\mathcal D}_{1,2}$
are very sharply distributed on the $Z_3$-axes,
and we can find three ``lines'' in the complex plain.
In any case, the use of ${\mathcal D}_{1,2}$ seems
much more advantageous than a Polyakov loop,
since it gives us more clearer signals.

\begin{figure}[h]
\includegraphics[scale=0.27]{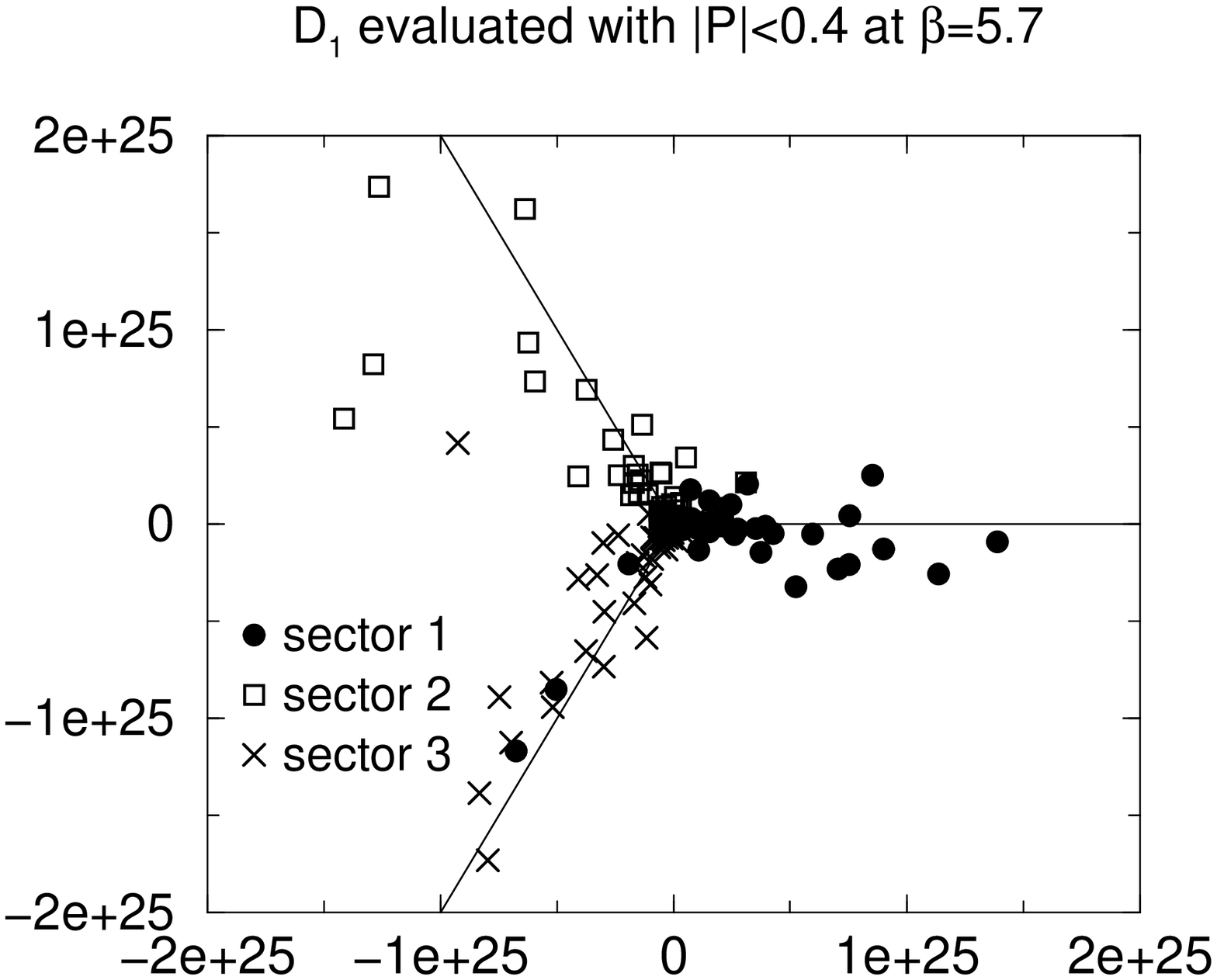}
\includegraphics[scale=0.27]{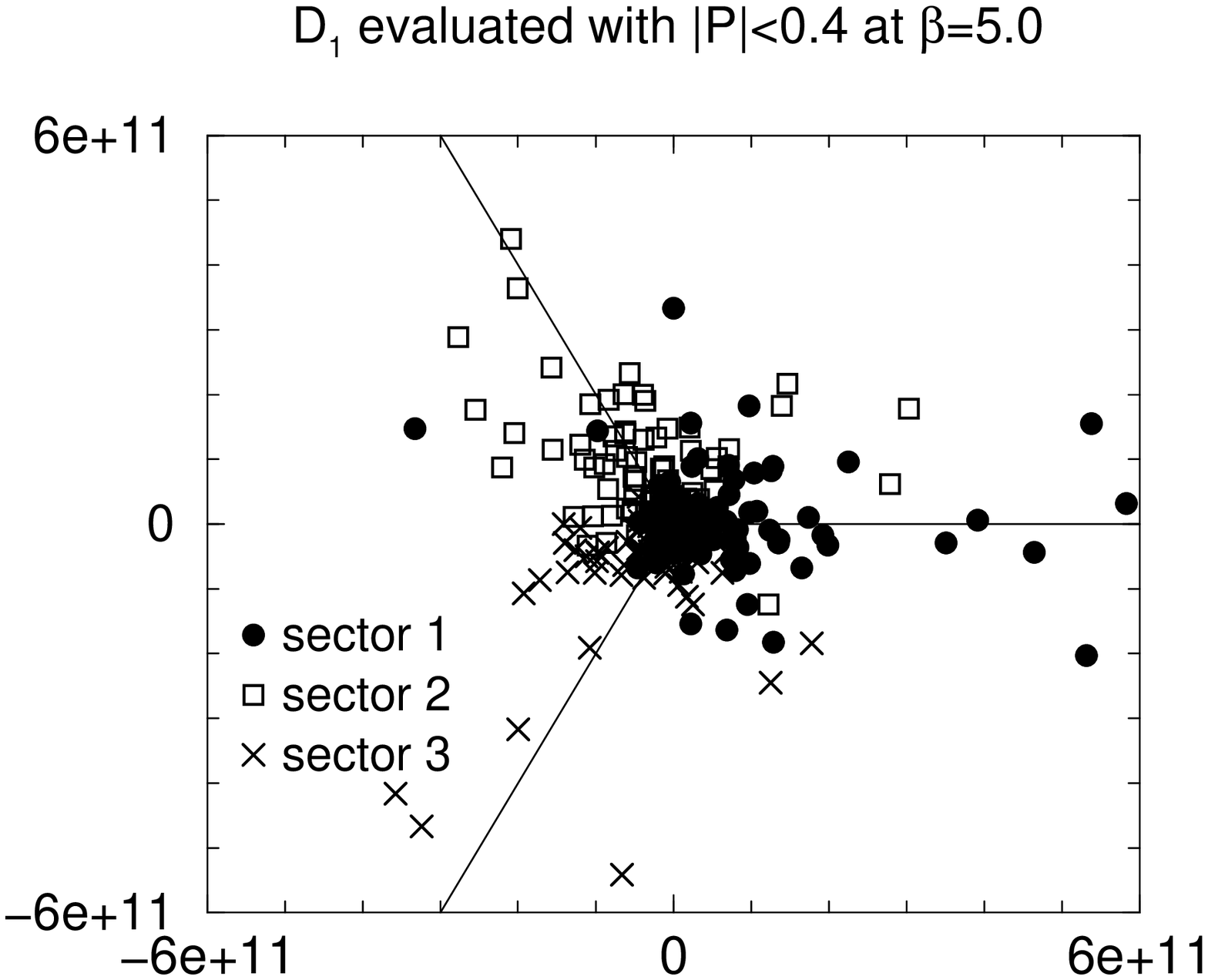}
\caption{\label{distFD04}
The distributions of ${\mathcal D}_1$
at $\beta = 5.7$(upper) and $\beta = 5.0$(lower),
obtained on the gauge configurations
where Polyakov-loop's absolute values are less than 0.4,
are plotted in the complex plain.
}
\end{figure}

The reason for the tiny deviations from the $Z_3$ axes
in Fig.~\ref{distFD}(upper)
is the discontinuous behavior in $D_{1,2}$ in the deconfined phase.
To see this, we compute ${\mathcal D}_1$
on the gauge configurations
where Polyakov-loop's absolute values are less than 0.4.
Such Polyakov-loop's distributions
are round-shaped both at $\beta =5.7$(deconfined) and 5.0(confined).
On the other hand,
${\mathcal D}_1$ at $\beta =5.7$ (Fig.~\ref{distFD04}(upper))
still reproduces the three-peak structure,
which implies ${\mathcal D}_1$ at small $|P|$ also reflects the phase.
To have a closer look,
we classify configurations into three categories,
sector 1 ($-\frac{\pi}{3} \leq {\rm arg} P \leq \frac{\pi}{3}$),
sector 2 ($\frac{\pi}{3} \leq {\rm arg} P \leq \pi$),
sector 3 ($-\pi \leq {\rm arg} P \leq -\frac{\pi}{3}$),
in terms of the phase angle (${\rm arg} P$) of a Polyakov loop.
The scatter plot of ${\mathcal D}_1$ in each sector 
can be found in Fig.~\ref{distFD04}.
Though, at $\beta =5.7$, the distribution of a Polyakov loop
($|P| < 0.4$) is round-shaped,
that of ${\mathcal D}_1$ is split onto $Z_3$ axes,
which are drawn as three solid lines in the figure.
This splitting indicates that
${\mathcal D}_{1,2}$ ``as a function of $P$''
quickly changes its value
at the boundaries between sectors
in the deconfinement phase
whereas they seem to be rather smooth functions in the confinement phase.
${\mathcal D}_{1,2}$ would be discontinuous 
at the boundaries in the thermodynamic limit,
which is considered as
the consequence of the spontaneous $Z_3$-symmetry breaking.
Even if the $Z_3$ symmetry 
in the $Z_3$-symmetrized canonical formulation
is always spontaneously broken
in the presence of dynamical quarks
and the transition is crossover,
it will be still reflected in ${\mathcal D}_{1,2}$.

\section{Summary}

We have considered prescriptions that 
are free from the direct charge-screening effects by quark loops
and enable us to clarify the confining nature of a vacuum.
We have tested two candidates for an order parameter,
a Polyakov loop ($P$) evaluated in zero-triality sector
and fermionic determinants (${\mathcal D}_{1,2}$)
with non-zero triality.
We have also individually investigated the distribution of a Polyakov loop
in each triality sector.
Especially, ${\mathcal D}_{1,2}$
has much smaller fluctuations in comparison with a Polyakov loop
in zero-triality sector,
and seems to well reflect the characteristic of a vacuum.
Such prescription could single out the confinement nature of a vacuum
properly and independently of the direct screening effects or
the string-breaking effects.

\section*{Acknowledgement}
The author thanks T.~Onogi for fruitful discussions,
and thanks O.~Borisenko, M.~Faber, P.~de~Forcrand,
K.~Fukushima, and C.~Gattringer for helpful comments.
The numerical calculations in this work were performed
with the computer facility at YITP, Kyoto University.
This work was supported by a Grant-in-Aid for
Scientific research by Monbu-Kagakusho (No. 20028006) and 
Yukawa International Program for Quark-Hadron Sciences (YIPQS).


\begin{thebibliography}{99}

\bibitem{Philipsen:2007rj}
  O.~Philipsen,
  Eur.\ Phys.\ J.\ ST {\bf 152} (2007) 29
  [arXiv:0708.1293 [hep-lat]].

\bibitem{Gattringer:2006ci}
  C.~Gattringer,
  Phys.\ Rev.\ Lett.\  {\bf 97} (2006) 032003
  [arXiv:hep-lat/0605018].

\bibitem{Bruckmann:2006kx}
  F.~Bruckmann, C.~Gattringer and C.~Hagen,
  Phys.\ Lett.\  B {\bf 647} (2007) 56
  [arXiv:hep-lat/0612020].

\bibitem{Nambu:1961tp}
  Y.~Nambu and G.~Jona-Lasinio,
  Phys.\ Rev.\  {\bf 122} (1961) 345.

\bibitem{Nambu:1961fr}
  Y.~Nambu and G.~Jona-Lasinio,
  Phys.\ Rev.\  {\bf 124} (1961) 246.
 
\bibitem{McLerran:2007qj}
  L.~McLerran and R.~D.~Pisarski,
  Nucl.\ Phys.\  A {\bf 796} (2007) 83
  [arXiv:0706.2191 [hep-ph]].

\bibitem{Miura:2008gd}
  K.~Miura and A.~Ohnishi,
  arXiv:0806.3357 [nucl-th].

\bibitem{Detar:1982wp}
  C.~E.~Detar and L.~D.~McLerran,
  Phys.\ Lett.\  B {\bf 119}, 171 (1982).

\bibitem{Fukushima:2002bk}
  K.~Fukushima,
  Annals Phys.\  {\bf 304} (2003) 72
  [arXiv:hep-ph/0204302].

\bibitem{Faber:1995up}
  M.~Faber, O.~Borisenko and G.~Zinovev,
  Nucl.\ Phys.\  B {\bf 444} (1995) 563
  [arXiv:hep-ph/9504264].

\bibitem{Kratochvila:2006jx}
  S.~Kratochvila and P.~de Forcrand,
  Phys.\ Rev.\  D {\bf 73} (2006) 114512
  [arXiv:hep-lat/0602005].

\bibitem{deForcrand:1999cy}
  P.~de Forcrand and V.~Laliena,
  Phys.\ Rev.\  D {\bf 61} (2000) 034502
  [arXiv:hep-lat/9907004].

\bibitem{Aarts:2001dz}
  G.~Aarts, O.~Kaczmarek, F.~Karsch and I.~O.~Stamatescu,
  Nucl.\ Phys.\ Proc.\ Suppl.\  {\bf 106} (2002) 456
  [arXiv:hep-lat/0110145].

\bibitem{Sasai:2003py}
  Y.~Sasai, A.~Nakamura and T.~Takaishi,
  Nucl.\ Phys.\ Proc.\ Suppl.\  {\bf 129} (2004) 539
  [arXiv:hep-lat/0310046].

\bibitem{Alexandru:2005ix}
  A.~Alexandru, M.~Faber, I.~Horvath and K.~F.~Liu,
  Phys.\ Rev.\  D {\bf 72} (2005) 114513
  [arXiv:hep-lat/0507020].

\end{thebibliography}
\end{document}